# DesignCon 2007

# Applications of Optimization Routines in Signal Integrity Analysis

Pat J. Zabinski, Mayo Clinic
zabinski.patrick@mayo.edu

Ben R. Buhrow, Mayo Clinic
buhrow.benjamin@mayo.edu

Barry K. Gilbert, Mayo Clinic
gilbert.barry@mayo.edu, 507-284-4056

Erik S. Daniel, Mayo Clinic

# Abstract

Signal integrity analysis often involves the development of design guidelines through manual manipulation of circuit parameters and judicious interpretation of results. Such an approach can result in significant effort and sub-optimal conclusions. Optimization routines have been well proven to aid analysis across a variety of common tasks. In addition, there are several non-traditional applications where optimization can be useful. This paper begins by describing the basics of optimization followed by two specific case studies where non-traditional optimization provides significant improvements in both analysis efficiency and channel performance.

# Author(s) Biography

Pat Zabinski earned a BSEE from St. Cloud State (St. Cloud, MN) and MSEE from University of North Dakota (Grand Forks, ND). He is currently a principal engineer at the Mayo Clinic where he focuses on signal integrity aspects of integrated circuits, packages, boards, and systems.

Ben Buhrow earned a BS in Physics from the University of Northern Iowa (Cedar Falls, IA). He is currently a project engineer at the Mayo Clinic where he focuses on signal integrity aspects and testing of integrated circuits, packages, boards, and systems.

Barry K. Gilbert received a BSEE from Purdue University (West Lafayette, IN) and a Ph.D. degree in physiology and biophysics from the University of Minnesota. He currently is Director of the Special-Purpose Processor Development Group at the Mayo Clinic, directing research efforts in high performance electronics and related areas.

Erik S. Daniel received a BA in physics and mathematics from Rice University (Houston, TX) in 1992 and a Ph.D. degree in solid state physics from the California Institute of Technology (Pasadena, CA) in 1997. He currently is Deputy Director of the Special-Purpose Processor Development Group, directing research efforts in high performance electronics and related areas.

# Introduction

Signal integrity (SI) engineering often involves an iterative process of refinement of models, conditions, and analyses. In many of these efforts, automated optimization routines can provide tremendous value in improved efficiency and accuracy.

One of the more common applications of optimization routines is to develop equivalent-circuit models of various passive interconnect [1] [2]. In these cases, the interconnect performance is measured in the lab in either time- or frequency-domains. Based on the measured performance and knowledge of the physical structure, a circuit topology is generated to approximate the performance. Optimization is then used to adjust the various circuit element values to best approximate the measured performance.

Beyond model generation, optimization routines can be valuable in peaking performance of complete links such as determining optimal termination resistance. However, there are several non-traditional applications of optimization routines that can be of similar value. For example, it is often possible to associate specific design features with cost and thereby optimizing cost versus performance trade-offs.

This paper describes the basic principles of optimization along with concepts of how optimization can be used in non-traditional ways. To aid in understanding some of the practical implementation issues, two analysis example are included.

There are numerous commercially-available simulation tools that can readily perform the described optimization. However, to provide the opportunity for the reader to duplicate the analyses and explore further possible applications, we opted to include example code for a tool commonly used in industry - Synopsys' HSPICE.[1]

# Optimization Basics

The primary value of optimization routines is that they provide an automated means of determining specific values that best meet a performance objective. Because the iterative process is automated, optimization saves time and effort involved in manually running a series of iterative simulations. In addition, optimization often leads to more accurate results due to its ability to fine-tune parameters and simultaneously deal with multiple variables.

Optimization routines take many forms, each of which has its own respective strengths and weaknesses [3]. However, they generally share the same basic iterative-process of estimating values for variables, comparing the results to a performance goal, establishing new estimated values, and continuing the process until the resulting performance reasonably matches the target performance.

---

[1] Synopsys, Inc. 700 East Middlefield Road, Mountain View, CA 94043. The authors and their employer do not endorse any commercial simulation tool.

## Basic Setup Parameters

Beyond the actual circuit of interest, optimization routines require three basic pieces of information: 1) list of the variables to be adjusted; 2) a metric to evaluate the circuit performance; and 3) optimization control options.

In HSPICE, the variables are identified in .PARAM statements, the performance metric is described in .MEASURE statements, and the control options are identified in .MODEL statements. To link all optimization-related elements together, some form of analysis statement (i.e., .DC, .AC, or .TRAN) is also needed. The basic form of these statements for a transient simulation is shown below.

```
.PARAM
+ p1=opt1(init, min, max)
+ p2=opt1(init, min, max)

.MEASURE TRAN meas1 <OPTIONS> GOAL=...

.MODEL our_opt OPT <OPTIONS>

.TRAN tstep tstop LIN SWEEP
+ OPTIMIZE=opt1 RESULTS=meas1 MODEL=our_opt
```

To conserve text and due to much existing documentation on these options, this paper assumes the reader is either well acquainted with or has ready access to information on HSPICE options. Accordingly, we describe the basic features of each statement.

In the .PARAM statement, P1 and P2 are the optimization parameters to be tuned (e.g., resistor value, line width). The *opt1* keyword is the optimization parameter reference name and must match the *opt1* keyword in the analysis statement (e.g., .TRAN). If multiple parameters are to be tuned, they all must be identified in the same .PARAM statement and use the same optimization parameter reference name. For each variable, an initial (INIT) value must be given as well as the range of possible values (MIN and MAX).

The .MEASURE statement identifies the performance metric and associated goal for the circuit. The performance metrics can be single point values (e.g., voltage at a specific time) or complex functions (e.g., relative match of two waveforms over time). As will be shown in greater detail later, the .MEASURE statement provides the path towards the greatest flexibility and utility in optimization. Note that the *meas1* keyword must match that in the analysis statement (i.e., .TRAN).

The .MODEL command initiates the optimization routine within HSPICE and provides user control over various options and settings. The *our_opt* keyword is the optimization reference name, and it must match the model option in the analysis statement (e.g., .TRAN).

The .TRAN analysis statement signifies a transient simulation is to be run, and it ties various optimization statements together. Note that the the *opt1*, *meas1*, and *our_opt* keywords included in the .TRAN analysis statement must match those in the .PARAM, .MEASURE, and .MODEL statements.

Assuming a successful simulation run, HSPICE provides the results in tabular form that includes both the final optimized value a normalized-sensitivity value for each variable. The normalized sensitivity indicates which variables have the most impact on circuit performance, where variables with lower sensitivity have smaller impact on performance than variables with higher sensitivity.

## Case Study: Multi-Drop Bus

A common SI effort involves the analysis of multi-drop busses. In the example depicted in Figure 1, a processor drives a common address bus to four memory chips. From a practical perspective, the printed wiring board (PWB) traces leading to the four memory chips typically form a primary path from the processor to the last memory chip along with trace-stubs down to the intermediate memory chips. The challenges with this circuit topology include termination of each interconnect path, potential over-loading of the driver, and constructive/destructive interference from multiple reflections.

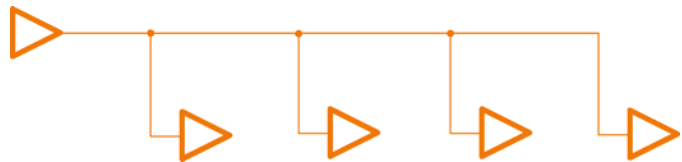

**Figure 1: Multi-Drop Bus Circuit Topology. (21742)**

A common approach to termination is to terminate the far end of the primary signal path with a 50 Ohm termination resistor to VTT. While such a termination scheme works well for simple point-to-point circuit topologies, the signal waveforms shown in Figure 2 clearly demonstrate the undesired effects of improper termination from the various stubs.

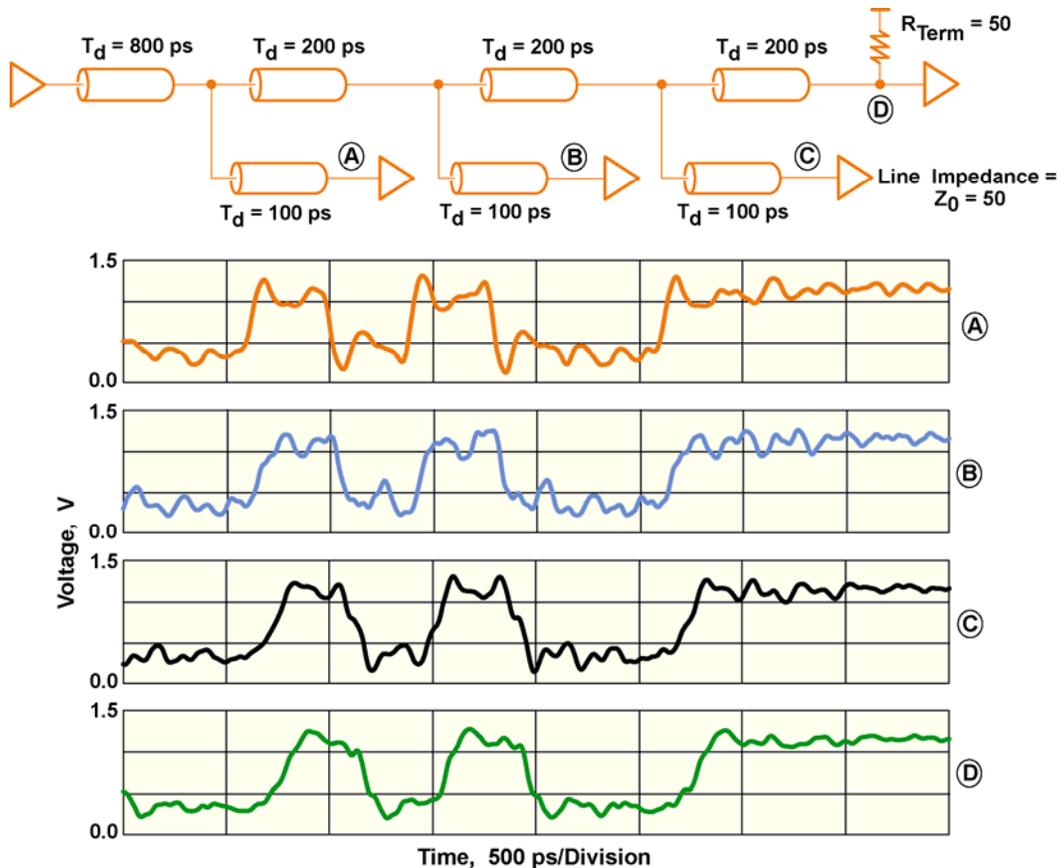
**Figure 2: Waveforms at the Four Loads Using Conventional Termination. (21743)**

Through manual inspection of these example waveforms, the signals collapse within 90 mV of the 750 mV reference voltage, which does not meet the 200 mV HSTL specification [4] for AC conditions. Accordingly, the design needs to be changed prior to relying on it in system applications.

## Quantifying Performance

Due to inter-symbol interference (ISI), the shape of each signal bit is influenced by its preceding and succeeding bits. Accordingly, it is usually prudent to consider more than a single bit in assessing the performance of a net that can have a variety of bit patterns. Instead, it is better to simulate the net with a wide array of bit patterns and analyze the waveform in its entirety using an eye diagram. The challenge with many simulation tools is that they do not directly measure eye openings, and thus some form of creative adaptation needs to occur. Specific to HSPICE, this adaptation occurs within .MEASURE and supporting statements.

When considering the signal voltage, $V_{SIG}$, it is often most useful to consider the voltage relative to a reference voltage, $V_{REF}$. Referring to Figure 3 (a), the signal margin of interest is represented by the small vertical arrows that indicate the distance from the $V_{REF}$ reference voltage and the $V_{SIG}$ signal. As an example, the HSTL standard [4] specifies this AC voltage margin should at least 200 mV beyond $V_{REF}$ in order to be considered a logic one or zero.

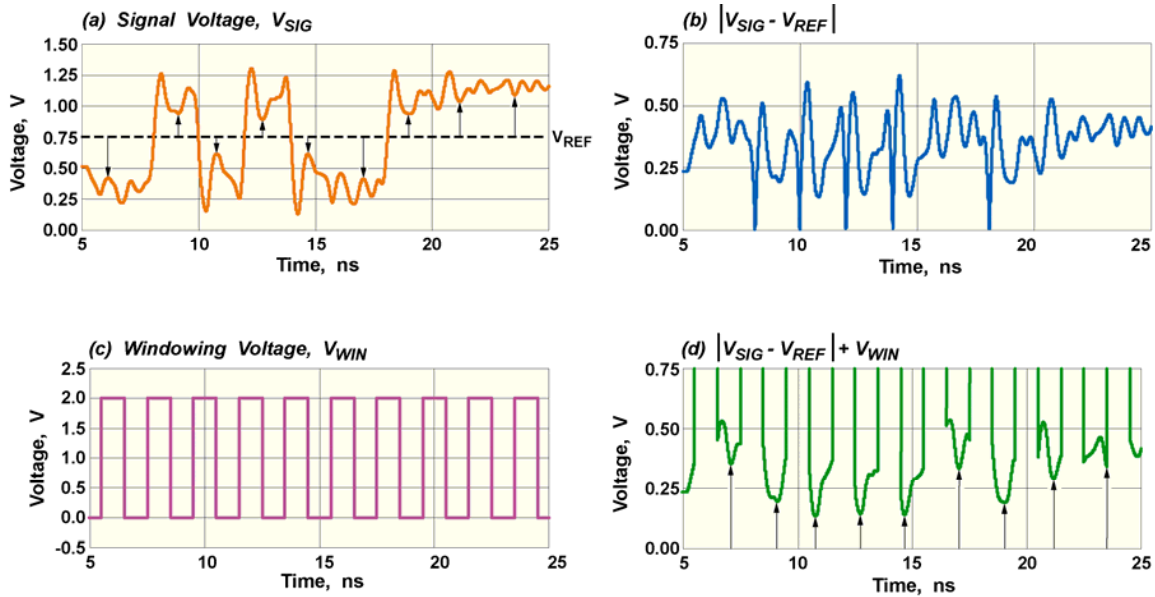

**Figure 3: Waveform Translations Used to Measure Eye Diagrams. (21777)**

Because the signal swings to both sides of the reference voltage, the voltage margin defined by $V_{SIG} - V_{REF}$ will have both positive and negative values, which makes it difficult to find one performance goal to use in optimization. To ignore the sign of the margin and focus on the magnitude, we can consider the absolute value of $V_{SIG} - V_{REF}$ as shown in Figure 3(b).

Considering Figure 3(b), a difficulty arises during the signal transition regions, shown in Figure 4, where the signal naturally and necessarily passes near $V_{REF}$. During these regions in time, $|V_{SIG} - V_{REF}|$ will always result in zero values, which disallows us from simply considering the minimal value over the entire time period. Instead, it is possible to ignore the transition region by adding a windowing waveform, $V_{WIN}$, per Figure 3(c) where a relatively large voltage is added to the $|V_{SIG} - V_{REF}|$ waveform during the transition regions, and nothing is added during the area of interest (i.e., open eye mask). The performance goal is now easily specified as maximizing the minimum value of $|V_{SIG} - V_{REF}| + V_{WIN}$ with 200 mV being the minimally-accepted value.

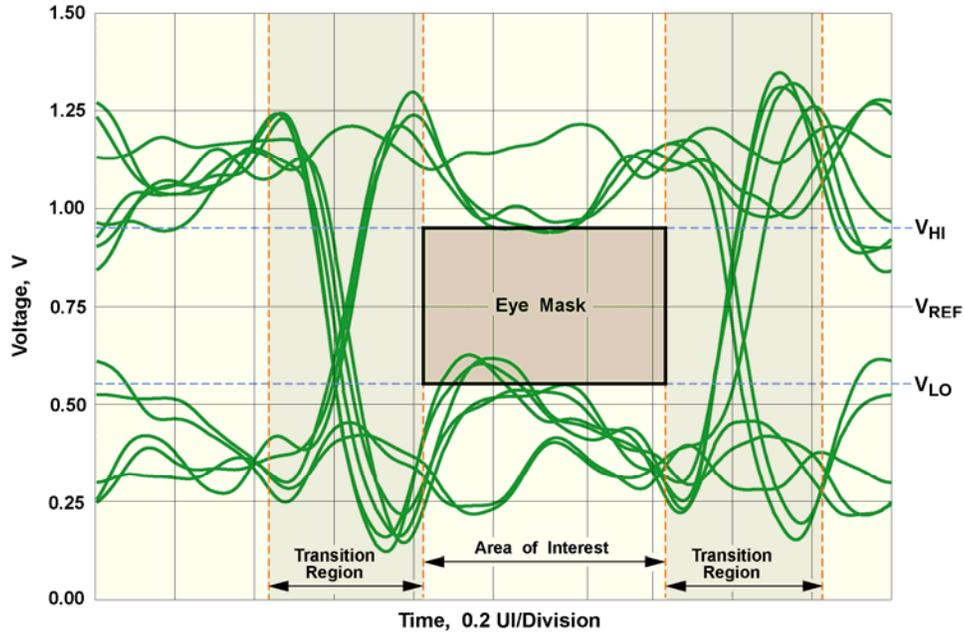
**Figure 4: Example HSTL Eye Diagram and Associated Mask. (21776)**

Figure 3(d) shows the final result of $|V_{SIG} - V_{REF}| + V_{WIN}$ where voltages during the transition periods are pushed well outside the range of interest. As a result, the lower portions of the signal represent the magnitude of the voltage difference between $V_{SIG}$ and $V_{REF}$ during the center portion of the eye. With this signal, the optimization then revolves around maximizing the minimal value across the entire bit stream, as highlighted by the small vertical arrows.

Here are the corresponding HSPICE statements needed to measure the effective eye opening.

```
.PARAM eye_mask=840ps
VWIN win 0 PULSE(2 0 delay 1f 1f 'eye_mask' period)
RWIN win 0 1meg
.MEAS TRAN eye_open MIN par('abs(v(sig)-vref) + v(win)')
```

where V(win) is the voltage used to window the eye during signal transitions, eye_mask is the eye's horizontal mask, delay is a tuning parameter to center the mask within the eye, and eye_open is the measured minimum eye opening (in voltage) over the entirety of the simulated waveform relative to vref.

Relating back to optimization, the goal is to maximize eye_open using any number of tools and tricks in the SI engineering toolbox.

## Optimization Process

A conventional application of optimization revolves around a process of selecting a particular circuit topology, optimizing the performance of that specific topology, and determining if that topology is sufficient for that application. If the optimized performance falls short of desired match to measurement, then another circuit topology is selected, and the process is repeated until a reasonable solution is found. Alternatively, optimization routines can be used to rapidly and simultaneously determine both the circuit topology and the associated parasitic values.

Considering the four-drop bus example, there are numerous termination options whereby a dozen termination resistors are placed in shunt and series at every major node, and the transmission line impedance is allowed to vary from the nominal 50 Ohm value. Such a thorough solution shown in Figure 5 is likely to provide good signal integrity; however, the expense of such a solution is impractical from perspectives of cost of the components, complexity of board design, and needed real estate for the components. Yet, by judicious use of optimization, it is often advantageous to start with such a "kitchen sink" option.

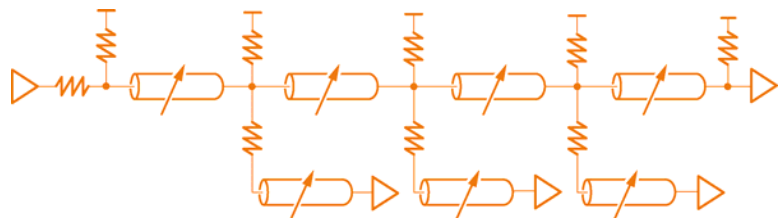

**Figure 5: "Kitchen Sink" Termination Options. (21747)**

More specifically, one often-useful application of optimization is to begin with a circuit topology that covers many practical options. Optimization is performed on the circuit, and the resulting parasitic values are inspected. If any parasitic value falls outside a useful range, the entire element is removed from the circuit, and optimization is performed again. The process continues until all parasitic values approach reasonable and practical values and the resulting signal integrity is adequate.

One common feature most optimization routines share is their increased inefficiency with an increase in the number of variables. If all twelve resistors and all seven line impedances in the above example were allowed to vary independently, the optimization process would take considerable time and the results could be impractical to implement. Alternatively, it is reasonable to constrain the primary path from the driver to the last receiver to one common line impedance while all the stubs rely on different impedance. Similarly, the three resistors at each transmission line split can be forced to be the same value for each split, thus further reducing the number of variables.

With most optimization routines, nominal and limit values are needed. For the transmission lines, 50 Ohm impedance is most-often used and will be the suggested nominal value for the simulations. For the limits, practical values are chosen to range from 25 Ohm to 100 Ohm. For the shunt resistors, nominal 50 Ohm values are chosen with limits of 1 Ohm and 1000 Ohm. For the series resistors, 10 Ohm nominal, 1 Ohm minimum, and 1000 Ohm maximum values are chosen. For all resistors, values approaching 1 Ohm will be considered shorts and values approach 1000 Ohm will be considered opens. In all, the nineteen possible variables are cut down to the eight listed below.

```
.PARAM
+ series_r_drvr=OPT1(10, 1, 1000)
+ series_r_primary=OPT1(10, 1, 1000)
+ series_r_stub=OPT1(10, 1, 1000)
+ shunt_r_drvr=OPT1(50, 1, 1000)
+ shunt_r_primary=OPT1(50, 1, 1000)
+ shunt_r_rcvr=OPT1(50, 1, 1000)
+ z_primary=OPT1(50, 25, 100)
+ z_stub=OPT1(50, 25, 100)
```

Using the circuit topology of Figure 5 and the parametric values shown above, optimization is run with the goal of each eye opening being maximized. Upon completion of optimization, HSPICE provides the following output.

|  | value | %norm-sen |
|---|---|---|
| .param series_r_drvr = | 1.0000 | 298.2954m |
| .param series_r_prima= | 4.2140 | 6.5805 |
| .param series_r_stub = | 13.9313 | 7.9376 |
| .param shunt_r_drvr = | 1.0000k | 3.6586 |
| .param shunt_r_primar= | 535.8809 | 10.2347 |
| .param shunt_r_rcvr = | 48.9906 | 25.5837 |
| .param z_primary   = | 61.2177 | 36.2221 |
| .param z_stub      = | 51.7493 | 9.4845 |

Considering the series resistor values, the series values for both the driver and the primary interconnect paths (series_r_drvr and series_r_primary) are both relatively low, and thus we consider them to be shorts and remove them from the circuit. Similarly, the shunt resistor values at the driver (shunt_r_drvr) and along the primary signal path (shunt_r_primary) are sufficiently high that we consider them to be opens, and thus remove them from the circuit as well. The resulting circuit is shown in Figure 6, which is much more practical to implement.

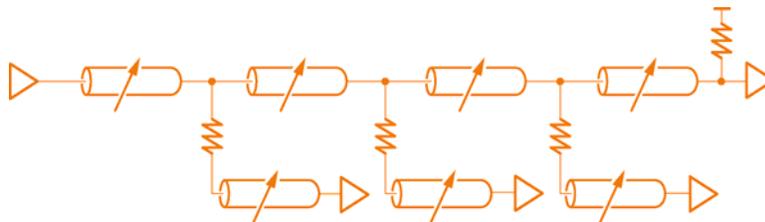

**Figure 6: Revised Termination With Removed Series and Shunt Resistors. (21748)**

Using the revised circuit, optimization is performed again to fine-tune the remaining parasitic values with these results.

|   | value | %norm-sen |
|---|---|---|
| .param series_r_stub = | 19.2114 | 8.0037 |
| .param shunt_r_rcvr = | 38.8936 | 38.1577 |
| .param z_primary = | 61.8679 | 47.6791 |
| .param z_stub = | 48.1519 | 6.1595 |

Considering these results, we round off the parasitics to practical values, and the resulting circuit performance is shown in Figure 7.

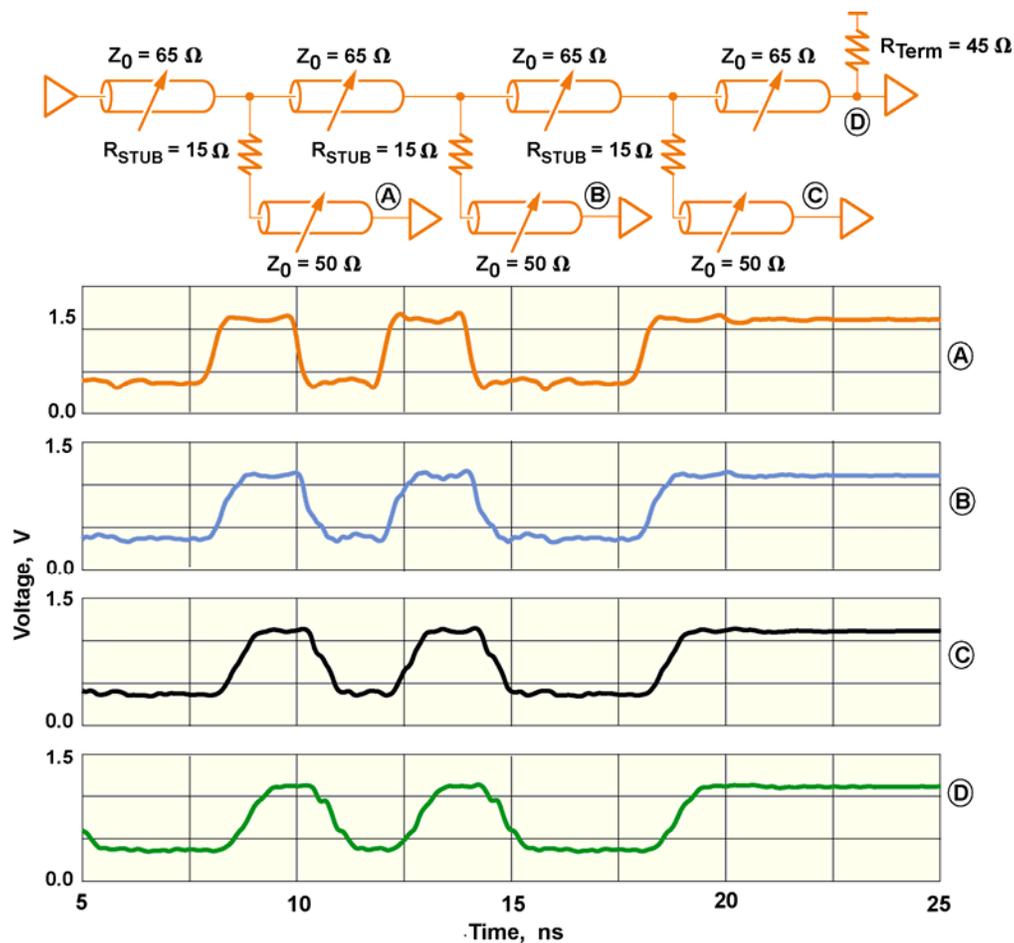

**Figure 7: Final Circuit Topology and Associated Waveforms for Multi-Drop Bus. (21749)**

### Results

The original non-optimized circuit provided an eye opening of just 90 mV beyond the reference voltage, produced poor signal waveforms, and failed to meet the HSTL specification. Through the demonstrated iterative optimization process, the eye opening improved to over 300 mV, the signal quality improved dramatically, and the signals now exceed the HSTL specification by 100 mV.

## Case Study: High-Speed Link

Another common SI effort involves the analysis of a high-speed serial data link on a PWB. The challenges with this type of analysis include signal degradation due to resistive loss, loss due to the dielectric material, reflections from vias and connectors, and crosstalk.

High-speed links are often required to travel large distances in the PWB, for instance to traverse a backplane from one daughter card to another. Because of these long distances, frequency-dependent loss is the dominating factor that must be overcome for good signal integrity at the receiver. Many commercial SerDes include pre-emphasis in transmitters and equalization in receivers to help compensate for these effects. Often, these features are the only ways to accommodate a given (long) length of interconnect in a high-speed system architecture.

The amount of loss present in the transmission line link is a consequence of the transmission line cross-sectional geometry and of the PWB material, both of which have practical constraints. For instance, while wider traces reduce losses, trace widths are often constrained to fit between vias under a ball grid-array (BGA) package. And there are obvious manufacturing constraints on the width of a trace. To better show this effect, consider the three simulated eye diagrams in Figure 8 that represent the same 1 m (40 inch) trace length but with different cross-sectional geometries for each. Pre-emphasis settings are the same for all three cases.

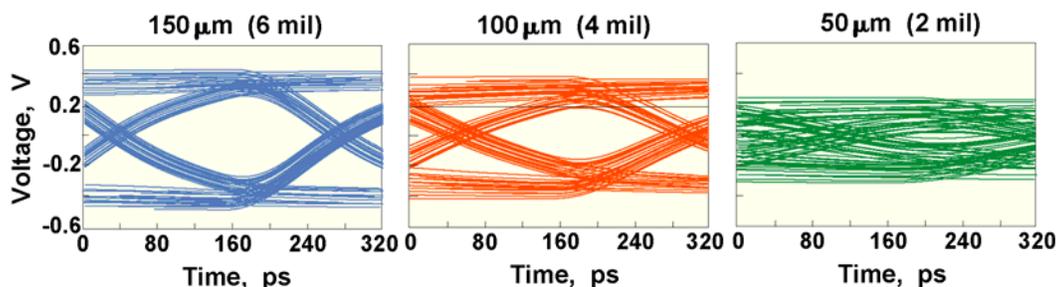

**Figure 8: Eye Diagrams After 1 meter (40") of PWB Differential Stripline Transmission Line at 6.25 Gbps. (21789)**

The plot corresponding to the 50 µm trace shows that resistive and dielectric losses are significant enough to completely close the eye. Alternatively, by widening the trace to 100 µm, the resistive losses are reduced to the point where the eye begins to open, and widening the trace even further to 150 µm yields further improvement. Note, however, that this trend does not continue indefinitely, because the dielectric loss starts to dominate over resistive loss, and no more benefit can be derived from increasing the trace width.

The design of the transmission line cross-section geometry is a complex task that impacts many factors throughout the system architecture. The overriding requirement, however, is an adequate eye diagram at the receiver. To achieve this requirement, it is often a good idea to err on the side of caution and simply use the widest trace physically possible to ensure the least amount of loss due to the interconnect.

However, the tradeoff for a wider trace is use of a thicker dielectric to maintain 100 Ohm differential impedance in the transmission line.  In contrast thinner dielectrics (and narrower traces) bring down board costs and reduce via geometries and thus their impact on electrical performance.  If the trace width can be reduced enough, perhaps entire signal layers can be removed because more routing channels between via fields can be opened up.  Thus, our goal should be the thinnest possible trace-width that maintains an adequate eye diagram.

This case study will go into the details of how we can use optimization to help with the cross-sectional geometry design of a typical high-speed link.  In this scenario, we seek to minimize the trace width for a variety of trace lengths.  This will create bounds and provide guidance on the cross-section design and maximum trace length for our link, and show that this process can improve upon simple conservative estimates. This methodology can be extended to other metrics, such as line width vs. pitch, or line length vs. dielectric loss tangent, but for this study we will concentrate on optimizing the line width.

## Quantifying Performance

As was the case for the multi-drop bus, we need a way to quantify the performance of our link over many bit periods.  Comparing an eye diagram to an eye mask is a natural way to do this, and is what is actually done in the laboratory on measured signals.  An eye diagram is just a super-position of many bits plotted on top of each other within a defined time base, where the natural time base is one bit period.  If we define our eye mask as a voltage vs. time waveform with one mask-pulse per bit period, then with the proper time alignment we can simply compare the signals on a point by point basis.

For high speed data links, eye masks are commonly given as irregular hexagonal shapes. We can easily create the top half of a hexagon using a PULSE voltage waveform in HSPICE.  For example, the following HSPICE statements create the top half of a hexagonal-shaped waveform where the maximum voltage is 0.165 V, the duration of the maximum voltage is 0.5 unit intervals (UI), and the total horizontal extent of the pulse shape is 0.6 UI.

```
.PARAM mask_rise='0.05 * bit_period'
.PARAM mask_fall='0.05 * bit_period'
.PARAM mask_high_time='0.5 * bit_period'

VMASK mask_p 0 PULSE(0 0.165 mask_delay mask_rise mask_fall
+ mask_high_time bit_period)
```

When the mask_delay value is chosen appropriately, the mask waveform is positioned in the middle of each bit period.  An eye diagram of the mask is shown in Figure 9, along with an associated waveform at 6.25 Gbps.  For clarity, we also plot the waveform of the inverse of the mask waveform, forming a complete hexagonal mask.  As will be shown later, this inverse mask waveform is not necessary in the actual analysis.

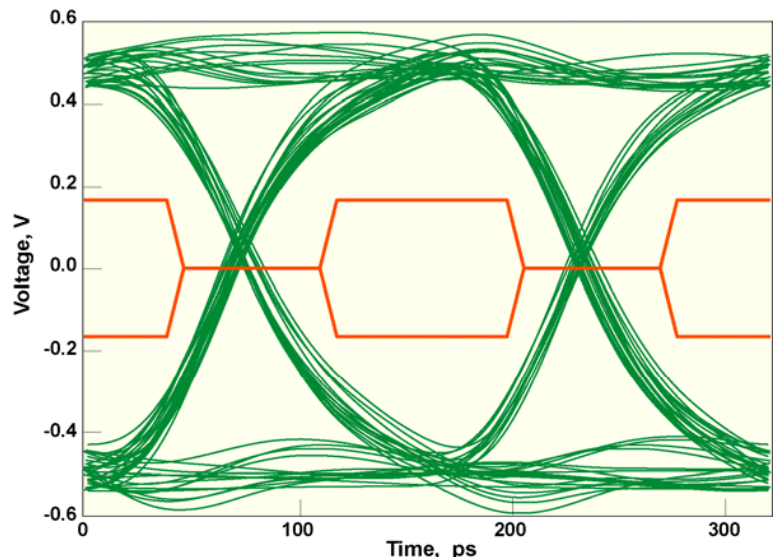
**Figure 9:  High-Speed Serial Data Signal With Mask Waveform.  (21790)**

The calculation of the appropriate delay setting for the mask is extremely important.  The best approach here differs quite a bit from the multi-drop bus case study because of the signal speed and because of the fact that we are trying to minimize the eye opening (subject to the masking constraint) rather than maximize it.  As we change the cross-sectional geometry, we also change the optimal delay setting for the mask relative to the signal.   In other words, as the signal degrades the best place for the eye mask may not be the exact center.  To deal with this anomaly, we will again turn to optimization, detailed in the next section.

Once we have the mask delay set, measuring the eye opening proceeds similar to the multi-drop bus case study.  Since we have differential signals that can take on both positive and negative voltage values, we use the absolute value function to fold the waveform over the crossing point voltage (zero volts) before subtracting the mask voltage, as shown in (a) and (b) of Figure 10.  We use a conditional statement in the .MEAS statement to ignore the transition regions by replacing the voltage value with a relatively high "out-of-bounds" value, as shown in (c) of Figure 10.  This scheme is analogous to the windowing function used in the multi-drop bus case study.

For instance, if we want to measure the minimum eye opening, the out-of-bounds value is chosen to be high, so that it does not enter into the MIN calculation of the .MEAS statement.  If we are interested in AVG or INTEGRAL, then we would choose an out-of-bounds value of 0.  Figure 10 shows the sequence of waveform transitions for the following MIN measurement:

```
.MEAS TRAN min_eye_opening MIN
+ par`((v(mask_p) > 0) ? (abs(v(inp,inn)) – v(mask_p)):10')
```

The exact minimum value of the MEAS statement can be read from the .mt0 file created by HSPICE and is available to the simulation engine during an optimization.

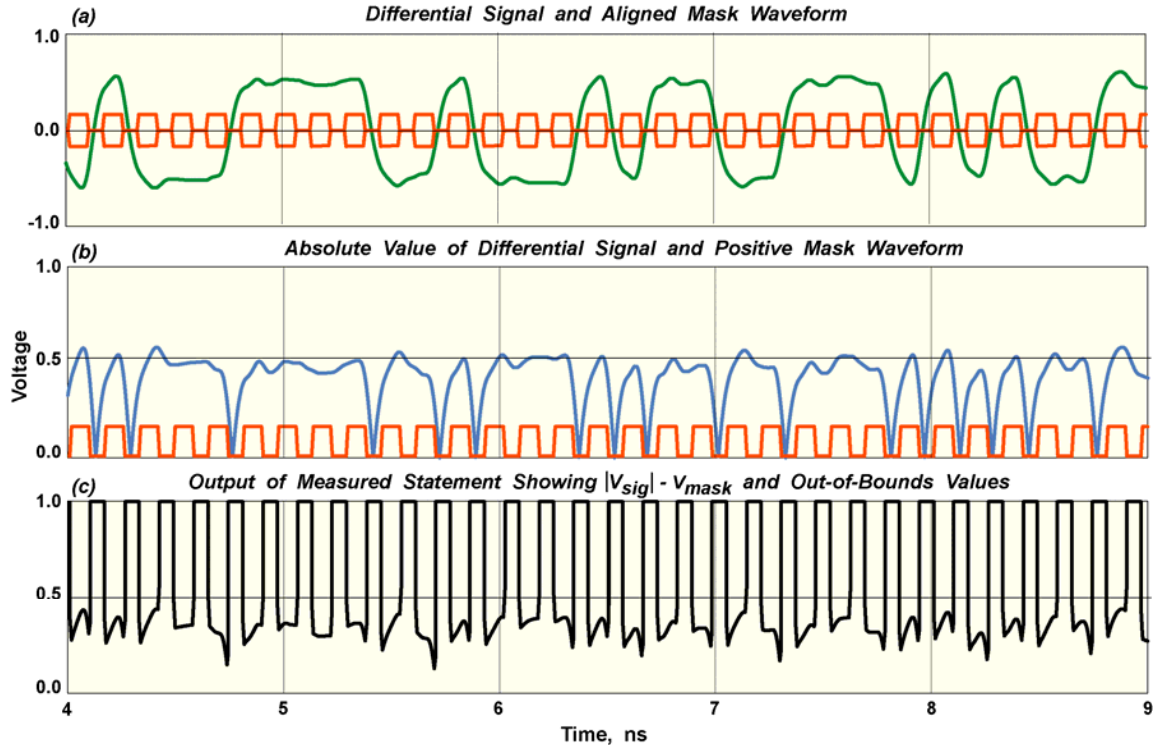

**Figure 10: Waveform Translations Used to Measure Eye Diagrams of High-Speed Serial Data. (21788)**

With the ability to measure the eye opening, we can now turn to our goal of minimizing the line width while maintaining an adequate quality eye diagram at the receiver.

## Optimization Process

The high speed serial data link we will work with is shown schematically in Figure 11; it consists of a length of differential transmission line representing a daughter card, connected to an identical daughter card through the same length of backplane and two identical connectors. The connectors are simple capacitor, transmission line, capacitor models that are not necessarily representations of real connectors; they simply represent some nominal impedance discontinuity. The data rate will be set to 6.25 Gbps.

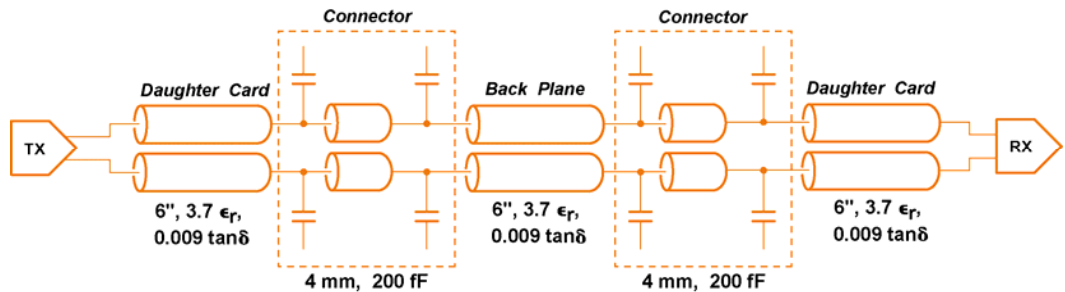

**Figure 11: Sketch of High Speed Serial Data Link. (21787)**

In order to simulate loss in the transmission lines, we must use a frequency-dependent model that takes into account both a finite metal conductivity and a non-unity dielectric permittivity. Also, in order to be able to change the model during an iterative optimization process, we must compute the model entirely within HSPICE. Fortunately, HSPICE has such a 2D modeling tool built in.

To generically define shapes and materials, HSPICE provides the .SHAPE and .MATERIAL statements. To define a specific PWB stackup, the .LAYERSTACK statement is provided. Finally, the .MODEL statement is used to specify a W-element model which points to the stackup and any options, and also defines the electrical conductors of the model. For example, to define a rectangular pair of conductors in a symmetric stripline type transmission line geometry one could use the following syntax:

```
.MATERIAL nelco DIELECTRIC ER=3.7 LOSSTANGENT=0.009
.MATERIAL cu METAL CONDUCTIVITY=57.6meg
.SHAPE rect RECTANGLE WIDTH=linewidth HEIGHT=metal_thickness

.LAYERSTACK stack_1
+ LAYER=(PEC,metal_thickness),
+ LAYER=(nelco,total_dielectric_t),
+ LAYER=(PEC,metal_thickness)

.MODEL w_diffpair W
+ MODELTYPE=FIELDSOLVER,
+ LAYERSTACK=stack_1,
+ FSOPTIONS=fs_options,
+ RLGCFILE=w_diffpair.rlgc,
+ CONDUCTOR=(SHAPE=rect, ORIGIN=(0,dielectric_t), MATERIAL=cu),
+ CONDUCTOR=(SHAPE=rect, ORIGIN=(pitch, dielectric_t), MATERIAL=cu)

.PARAM linewidth=101.6um
.PARAM metal_thickness=8.89um
.PARAM dielectric_t=122um
.PARAM total_dielectric_t=`dielectric_t * 2 + metal_thickness'
.PARAM pitch=254um
```

Parameters defined in the .LAYERSTACK or .MODEL statements can be used as inputs to the optimization routine just like any other parameter. If the value changes in the course of the optimization, HSPICE will automatically recalculate the W-element model with the new dimensions. The results of all model calculations are output to the file w_diffpair.rlgc, and are appended as necessary. For example, an optimization parameter statement we could use is:

    .PARAM linewidth=OPT1(101.6um, 50.8um, 127um)

In this case, we pick a conservative starting value that we believe will allow the link to meet our eye requirements at the receiver. The minimum value is optimistically low, with the goal that the optimization routine will drive the line width as close as possible to 50.8 μm while maintaining an adequate eye diagram at the receiver.

We must also deal with the fact that if we were to only change the line width, the transmission line differential-impedance would deviate from 100 Ohm. This might cause the resulting eye diagram to close sooner than it would if we were simply increasing the loss in the transmission line, due to reflections between the 100 Ohm driver and receiver and the transmission line.

We get around this obstacle by observing that if the metal thickness is small compared to the dielectric thickness, then the dielectric thickness and trace pitch should scale roughly linearly with the line width in order to maintain a 100 Ohm impedance. This observation is only valid within a certain range of line width values, outside of which our assumptions about metal thickness breaks down. However, a few simple simulations with a 2D EM solver reveal that this linear relationship is within 5% of the nominal characteristic impedance of 100 Ohm over the range of line width values we have chosen in the .PARAM statement above, compared to ~36% if we only change the line width. The 5% deviation occurs at the lower limit of 50 μm line width. Increasing the line width accumulates error even more slowly, leading to the conclusion that this simple scaling factor would work very well for other designs using the same metal thickness for signal traces. Implementing the scaling factor in HSPICE is pretty straightforward:

    .PARAM scale_factor=`linewidth/nom_linewidth'
    .PARAM pitch=`scale_factor * nom_pitch
    .PARAM dielectric_t=`scale_factor * nom_dielectric_t'

We define a nominal set of parameter values which have been pre-calculated to yield 100 Ohm impedance, and scale the nominal values according to the ratio of the current line width to the nominal line width.

In our process, the optimization routine will change the value of the parameter linewidth, which in turn will scale the pitch and dielectric_t parameters, and the new W-element model that is subsequently created will still have very near to 100 Ohm characteristic impedance. Any eye closure due to a change in line width will be a result of increased series resistance, and not due to stray reflections from unexpected impedance discontinuities.

Our primary optimization routine is the minimization of the line width. However as we mentioned before, the position of the mask relative to the signal will also need to be optimized because as the line width is changed, the shape of the signal's waveform also changes. Unfortunately, the goals of these two tasks directly conflict with each other. To optimize the mask position, the goal is the largest separation between mask and signal possible, while line width optimization's goal is the smallest separation. Rather than use a single optimization to try to optimize two conflicting goals (which we found does not work very well in practice), we chose to use several independent optimization statements run sequentially.

We found that four rounds of optimization are sufficient to find nearly the absolute minimum line width for a given length. Figure 12 illustrates this process. The first round, as shown in (a), finds the optimal mask delay in order to position the mask in the best location inside the eye. A second optimization (b) minimizes the line width for that specific mask. Almost always, one edge of the signal or the other causes the optimization to finish, but not both at the same time. For instance, in Figure 12 it is the falling edge that touches the lower bound of the mask. This means that if we were to adjust the mask delay again, line width minimization could continue. So, the third round (c) of optimization again adjusts the mask delay and the fourth (d) performs the final line width minimization.

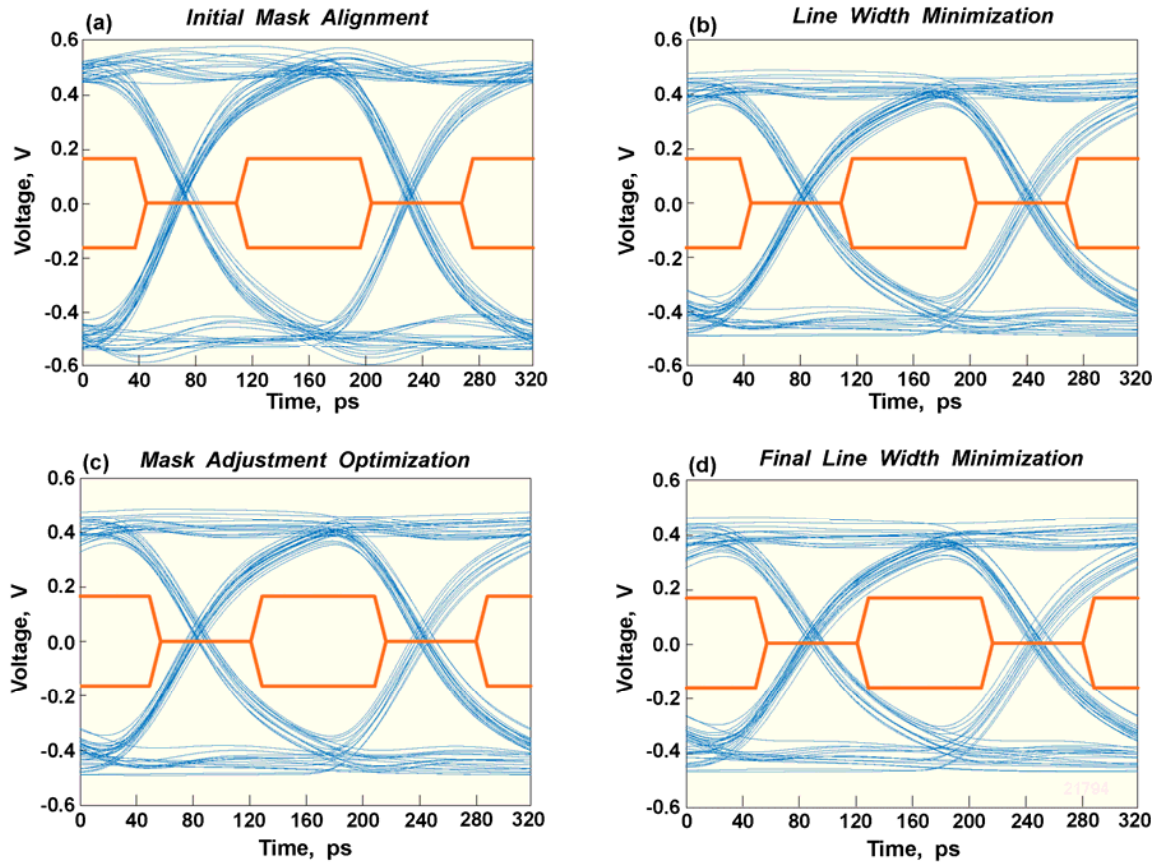

**Figure 12: Optimization Progression of Line-Width Minimization. (21794)**

The generic HSPICE syntax for running four separate optimizations sequentially in one netlist is as follows:

```
.PARAM param_1=OPT1(start, min, max)
+ param_2=OPT2(start, min, max)
+ param_3=OPT3(start, min, max)
+ param_4=OPT4(start, min, max)

.TRAN sweep_step sweep_stop
+ OPTIMIZIE=opt1 RESULTS=meas1 MODEL=model1
.TRAN sweep_step sweep_stop
+ OPTIMIZIE=opt2 RESULTS=meas2 MODEL=model2
.TRAN sweep_step sweep_stop
+ OPTIMIZIE=opt3 RESULTS=meas3 MODEL=model3
.TRAN sweep_step sweep_stop
+ OPTIMIZIE=opt4 RESULTS=meas4 MODEL=model4
```

The first transient analysis will run an optimization on param_1. Then the second transient analysis will run an optimization on param_2, using the final result of param_1 from the previous optimization, and so on. The key for this to work is that HSPICE carries over the results from one optimization to the next.

When we are trying to minimize the line width, we use the following measure statement to measure the eye opening:

> .MEAS TRAN min_eye_opening MIN
> + par`((v(mask_p) > 0) ? (abs(v(inp,inn)) – v(mask_p)):10')

The use of such statements was explained in detail in the Quantifying Performance section. For the purposes of optimization, we add GOAL=1E-5 to the measure statement because our goal is to make the difference between signal and mask as small as possible without causing mask violations. Note that we do not use a goal of exactly zero, because of the function HSPICE uses internally to evaluate the error between the current measurement and the goal would be undefined.

$$ERRfun = (GOAL - result)/GOAL$$

In this equation 'result' is the value returned by the .MEAS function, in this case the variable min_eye_opening. As we would like to avoid division by zero, we instead use a small value close to zero. This causes ERRfun to be large, but not infinite, and the optimization routine can still find a minimum of the function over the parameter space of interest.

When we are trying to maximize the eye opening while adjusting the mask delay, we use a similar measure statement except we add GOAL=1, and use the AVG function instead of MIN. The average difference is on the order of hundreds of millivolts, so a goal of one volt simply tells the optimization to make it as big as possible.

Any of several functions would work for this purpose, including MIN and INTEG (the area between the mask and the signal waveforms), but AVG was chosen because it is perhaps the most intuitive. An illustrative graphical representation of this eye mask optimization appears in Figure 13. In it, we are simply graphing the output of the measure statement as we sweep the mask delay parameter over two bit periods. When the average margin is relatively big then we know that the mask delay is positioned in such a way as to give an open eye. The output of the measure statement is periodic with respect to the mask delay parameter, with the same period as the signal bit period. The goal of the optimization is to find the mask delay that gives one of these maxima.

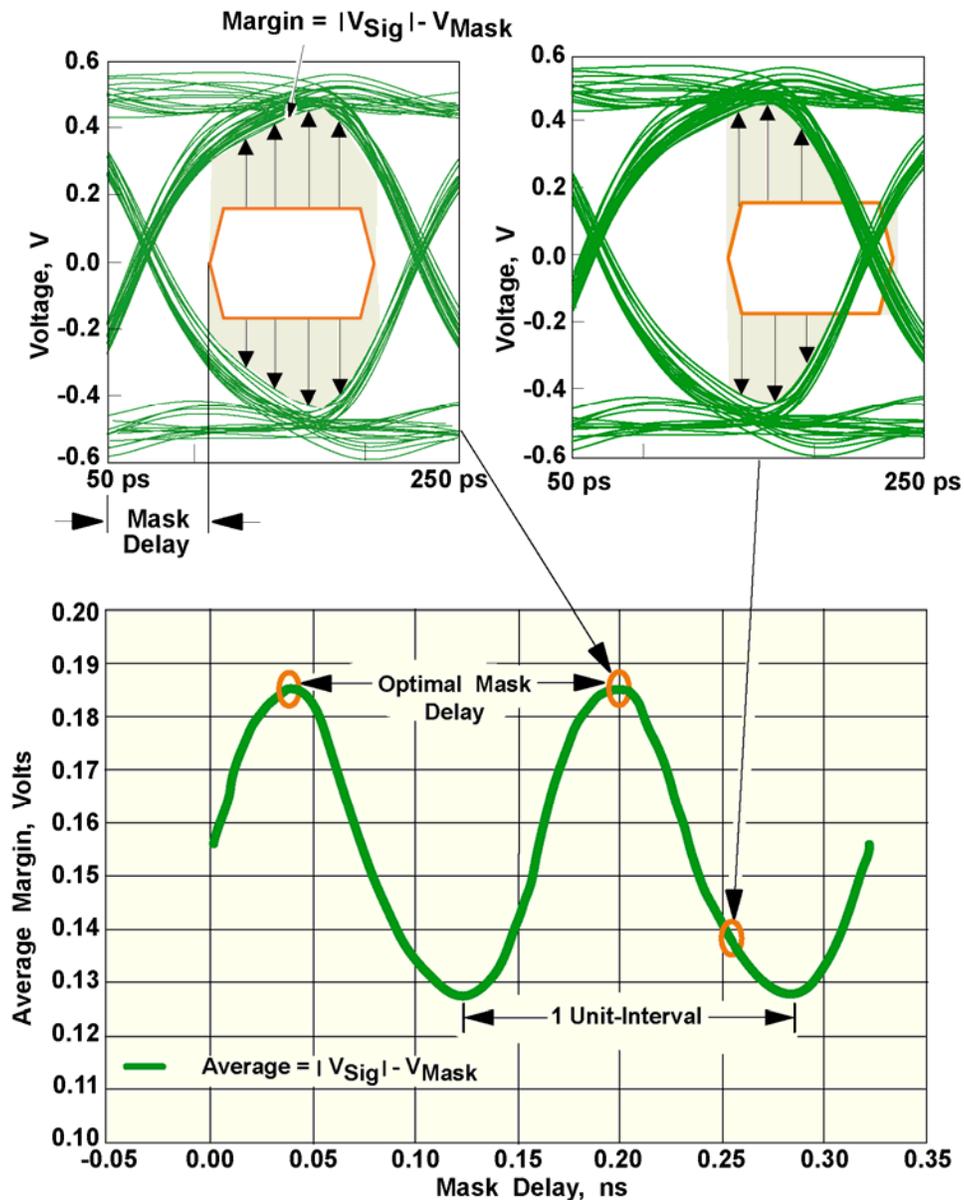

**Figure 13: Sweep of Parameter 'Mask_Delay' While Measuring Average Eye Opening. (21791)**

## Results

We ran our optimization sequence on line lengths ranging from 150 mm to 600 mm. We would expect to be able to use a narrower line width as the line length is reduced, because we can tolerate more loss. The actual observed relationship is shown in Figure 14. For long lengths, i.e. greater than 500 mm, the line width is actually increased from the starting value of 101.6μm by the optimization. Less than 500 mm, and the minimum tolerable line width also decreases, often by significant amounts. Of course, in conventional (e.g. laminate) technologies, it would be impractical to actually fabricate a stripline transmission line only 25 μm wide. The results show that for shorter lengths, the line width can be driven to the minimum feasible value. The plateau at the longer line widths represents the fact that the line width was capped at 121µm by the optimization routine.

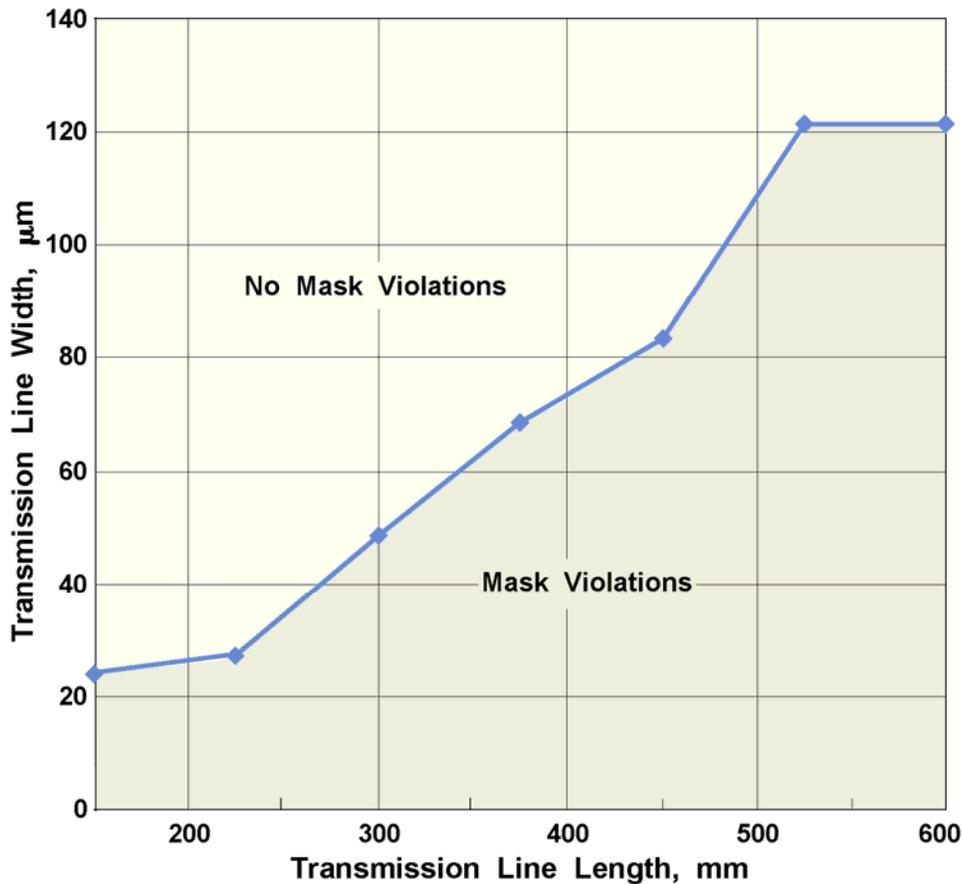

**Figure 14: Trace Length Versus Optimized Trace Width. (21793)**

This plot and the routines that generated it are useful in a number of ways. If the board size and hence the dielectric thickness are constrained, then we can quickly determine the maximum trace length possible for that board. For a given length we can determine the optimum trace width, and perhaps make board routing easier by using narrower transmission lines. Finally for a given length and width, we can quickly see if the eye will be open or have mask violations at the receiver. Once the optimization process is understood, it is a simple matter to apply it to other PWB geometries with different I/O models. There is an additional need to validate results, but the optimization process itself can save significant engineering time in system and board level design.

It should be mentioned that our final optimization process was not as robust as it was hoped it could be. In particular, optimization of the mask delay turned out to be very sensitive to the starting value, and often would not converge to the maximum shown in Figure 13 unless a mask delay value yielding a voltage margin close to a true maximum value was chosen. Also, ideally, we would like to iterate the process illustrated by Figure 12 until the mask delay adjustment converges to zero. This also was not possible;

therefore we chose to implement a fixed number of rounds of this iteration. A possible solution, barring changes to the tool used, would be a combination of optimization in HSPICE with external scripts. These scripts would parse the output, modify the optimization parameters, and run in a loop to convergence.

## Additional Considerations

Although optimization can be a powerful tool, there are plenty of limitations and drawbacks to consider, one of which is the amount of time needed to master proper use of the tool. Although most tools tend to come with supporting documentation, nothing is more valuable than hands-on experience. Prior to relying on optimization, it is useful to experiment with simple circuits whereby many optimization options can be explored and characterized. Specific to HSPICE, we found the .MODEL options CLOSE, RELIN, and RELOUT to be quite useful in helping with convergence and accuracy.

Also, optimization is not guaranteed to produce optimal or useful results. Through the course of preparing for this paper, the authors ran into several situations where the optimization routine produced unsuitable results. We recommend validating each result by viewing the associated circuit performance (e.g., eye diagram) and reviewing the optimization parameter output to validate convergence.

Many optimization runs can require a large number (e.g., tens to hundreds) of circuit simulations to converge to an adequate result. As a result, overall simulation times can be quite lengthy. As a first step towards easing the associated pain of long simulations, we recommend simplifying the circuit as much as possible.

As a second step towards improving simulation time, dramatic improvements can sometimes be made by replacing active circuit models with behavioral models. For example, line drivers (a.k.a., transmitters) can often be replaced with piece-wise linear voltage sources. In HSPICE, the .STIM command readily creates a piece-wise linear source that accurately emulates the performance of a device-level model when the loads are approximately the same. Similarly, the active circuitry of a receiver is generally not so much as important as the packaging and termination impedances, thus the receiver model can often be simplified with a handful of discrete parasitics.

Whenever making such simplifications, it is important to validate that the simplified models adequately emulate the device-level models under all practical conditions. As such, we recommend generating PWL models under various conditions and comparing the resulting models relative to one another prior to using any of the models in optimization.

## Conclusions

As demonstrated in the two examples, judicious use of optimization routines can save significant effort due to optimization's automated means of determining appropriate values to meet specific performance goals, be that goal peaked performance or minimal performance subject to constraints. In addition, optimization can be of tremendous help in determining appropriate circuit topologies for both models and circuits, and can often be used to navigate through the various conflicting goals of cost, reliability, and performance.

Yet, optimization should be used with caution, because issues such as non-convergence and inappropriate results can occur often. Therefore, it is imperative that all results are validated, and it is recommended that users familiarize themselves with all optimization options prior to proceeding.